\documentclass[letterpaper,english,aps,prl,twocolumn,floatfix,showpacs,amsfonts,amssymb,superscriptaddress]{revtex4}

\usepackage[T1]{fontenc}
\usepackage[latin1]{inputenc}
\usepackage{graphics}
\usepackage{amssymb}
\usepackage{amsmath}
\usepackage{overpic}

\newcommand{\beq}{\begin{equation}}
\newcommand{\eeq}{\end{equation}}
\newcommand{\bea}{\begin{eqnarray}}
\newcommand{\eea}{\end{eqnarray}}

\newcommand{\rmd}{{\rm d}}
\newcommand{\rmi}{{\rm i}}

\newcommand{\cuoo}{CuO$_{2}$}
\newcommand{\lco}{La$_{2}$CuO$_{4}$}

\begin{document}

\title{One-magnon Raman scattering in {\lco}: the origin of the field-induced mode}

\author{M.~B.~Silva~Neto}

\email{m.barbosadasilvaneto@phys.uu.nl}

\affiliation {Institute for Theoretical Physics, University of Utrecht,
  P.O. Box 80.195, 3508 TD, Utrecht, The Netherlands}

\author{L.~Benfatto}

\affiliation
{SMC-INFM and Department of Physics, University of Rome ``La
  Sapienza'',\\ Piazzale Aldo Moro 5, 00185, Rome, Italy}

\date{\today}

\begin{abstract}

We investigate the one-magnon Raman scattering in the layered
antiferromagnetic {\lco} compound. We find that the Raman signal is
composed by two one-magnon peaks: one in the $B_{1g}$ channel,
corresponding to the Dzyaloshinskii-Moryia (DM) mode, and another in the
$B_{3g}$ channel, corresponding to the XY mode. Furthermore, we show that a
peak corresponding to the XY mode can be induced in the planar $(RR)$
geometry when a magnetic field is applied along the easy axis for the
sublattice magnetization. The appearance of such field-induced mode (FIM)
signals the existence of a {\it new magnetic state} above the N\'eel
temperature $T_{N}$, where the direction of the weak-ferromagnetic
moment (WFM) lies within the {\cuoo} planes.

\end{abstract}


\maketitle

The possibility of inelastic light scattering by one- and two-magnon 
excitations in magnetic insulators was acknowledged long ago by Elliot 
and Loudon \cite{Elliot-Loudon} and the microscopic theory for such 
process was described in detail by Fleury and Loudon \cite{Fleury-Loudon}. 
One of the possible mechanisms for the magneto-optical scattering is 
an indirect electric-dipole (ED) coupling via the spin-orbit interaction,
and such mechanism has been used to determine the spectrum of magnetic 
excitations in many different condensed matter systems like fluorides, 
XF$_{2}$, where X is Mn$^{2+}$, Fe$^{2+}$, or Co$^{2+}$ \cite{Fleury-Loudon}, 
inorganic spin-Pierls compounds, CuGeO$_{3}$ \cite{Spin-Pierls}, and 
the parent compounds of the high-temperature superconductors, such as 
{\lco} \cite{Gozar}.

In this communication we report on a theoretical study of the Raman
spectrum in a two-dimensional (2D) quantum Heisenberg antiferromagnet
(QHAF) with DM and XY interactions, which is a model for the magnetism of a
single layer in {\lco}. As it was recently shown in \cite{MLVC}, although
small, such anisotropic DM and XY terms are very important in order to
explain the unusual magnetic-susceptibility anisotropies observed in {\lco}
\cite{Ando-Mag-Anisotropy}. It was found, in particular, that, below the
N\'eel ordering temperature, $T_{N}$, the two transverse magnon modes are 
gaped, with their respective gaps being uniquely determined by the 
strength of the DM and XY interaction terms, in agreement with neutron
scattering experiments \cite{Keimer}. However, more recent Raman-spectroscopy
experiments reported on the presence of {\it only one} of the above 
two magnon modes, the DM mode, in the Raman spectrum for the planar 
$(RL)$ geometry ($R$=right-rotating and $L$=left-rotating) at zero 
magnetic field \cite{Gozar}. Most surprisingly, it was also found that 
a second mode can be induced in the {\it forbidden} $(RR)$ geometry, 
for a magnetic field applied along the easy axis for the staggered 
magnetization, and only for such configuration. As we now clarify, 
the appearance of such FIM is directly associated to a continuous 
rotation of the spin quantization basis, which modifies the selection 
rules set by the ED coupling on the Raman scattering cross-section, 
thus allowing for the observation of the, previously selected out, 
XY magnon mode in the spectrum. 

The Hamiltonian representing the interaction of light with
magnons can be written quite generally as \cite{Cottam-Lockwood}
\beq
H_{ED}=\sum_{{\bf r}}{\bf E}_S^T\chi({\bf r}){\bf E}_I,
\eeq
where ${\bf E}_S$ and ${\bf E}_I$ are the electric fields of the
scattered and incident radiation, respectively (${\bf a}^T$ is the
transposed of the ${\bf a}$ vector) and $\chi({\bf r})$ 
is the spin dependent susceptibility tensor. We can expand  
$\chi({\bf r})$ in powers of the spin-operators, ${\bf S}$, as
\bea
\chi^{\alpha\beta}({\bf r})=\chi^{\alpha\beta}_0({\bf r})&+&
\sum_\mu K_{\alpha\beta\mu}S^\mu({\bf r})\nonumber\\
&+&
\sum_{\mu\nu} G_{\alpha\beta\mu\nu}
S^\mu({\bf r}) S^\nu({\bf r})
+\dots,
\eea
where $\mu,\nu=x,y,z$ label the spin components. The lowest order 
term $\chi^{\alpha\beta}_0({\bf r})$ is just the susceptibility 
in the absence of any magnetic excitation (it corresponds to elastic 
scattering), and it will be neglected in what follows. The second and 
third terms can give rise to one-magnon excitations because they can 
be written as $S^\pm({\bf r})$ and $S^z({\bf r})S^\pm({\bf r})$
respectively. The intensity of the scattering, as well as the selection 
rules, will be determined by the structure of the complex tensors 
$ K$ and $G$. For a square-lattice antiferromagnet the 
linear term on the spins reduces to (for Stokes scattering only)
\bea
H_{ED}^{AS}={\rmi} K_0\sum_{i_A,i_B}&&[(E_S^zE_I^+-E_S^+E_I^z)S_{i_A}^-
\nonumber\\
&&-(E_S^zE_I^--E_S^-E_I^z)S_{i_B}^+],
\label{AS-Hamiltonian-1}
\eea
where the sum runs over the two sub-lattices $A$ and $B$, the $K_0$ coupling
constant is the same for the two sub-lattices, and $z$ is the direction of
the spin easy-axis \cite{Cottam-Lockwood}. Analogously, the quadratic term
on the spins reduces to
\bea
H_{ED}^S=G_0\sum_{i_A,i_B}& &[(E_S^zE_I^++E_S^+E_I^z)O_{i_A}^-
\nonumber\\
& &+(E_S^zE_I^-+E_S^-E_I^z)O_{i_B}^+],
\label{S-Hamiltonian-1}
\eea
where $G_0$ is a coupling constant and $O_{i}^{\pm}=S_i^zS_i^\pm+S_i^\pm S_i^z$
\cite{Cottam-Lockwood}. We see that while the ${ K}$ tensor is
purely imaginary and totally anti-symmetric, the ${ G}$ tensor
is purely real and totally symmetric, as required by the general
symmetry properties of the electric susceptibility of a magnetic
material, also known as Onsager relations \cite{Landau-Lifshitz}. 
Furthermore, the relative values of the coupling coefficients, $K_0$ 
and $G_0$, can be deduced from the measurements of magneto-optical 
effects like: magnetic circular birefringence (to determine $K_0$) 
and magnetic linear birefringence (to determine $G_0$) 
\cite{Cottam-Lockwood}. 

The total ED Hamiltonian $H_{ED}=H_{ED}^{AS}+H_{ED}^{S}$ can be finally
written in terms of the $x,y,z$ components of the sublattice magnetization,
${\bf M}_i=({\bf S}_{i_A}-{\bf S}_{i_B})/2$, as 
\beq
H_{ED}=\sum_{i}\left\{ {\bf E}_S^T \chi^x {\bf E}_I{\;}M^x_i
+ {\bf E}_S^T \chi^y {\bf E}_I{\;}M^y_i\right\},
\label{H-ED}
\eeq
where we introduced the matrices
\bea
\chi^x{=}
  \left( \begin{array}{ccc}
     0 & 0 & d \\
     0 & 0 & 0 \\
     d^* & 0 & 0
         \end{array} \right)\!\!,{\;\;\;\;\;}
\chi^y{=}
  \left( \begin{array}{ccc}
     0 & 0 & 0 \\
     0 & 0 & d \\
     0 & d^* & 0
         \end{array} \right)\!\!,
\nonumber
\eea
with $d=G_0 S + \rmi K_0$. Here we made the usual mean-field assumption
$\langle S^z_{i_A}\rangle=-\langle S^z_{i_B}\rangle=-S$ and we dropped
terms of the type $S_{i_A}^{x,y}+S_{i_B}^{x,y}$ since these will give rise
to negligible contribution for the ${\bf k}=0$ scattering in the 
long-wavelength limit. In deriving Eq.\ (\ref{H-ED}) we assumed that the
spins order along the $z$ direction. In the more general case, the
spin easy-axis is along a different direction $\tilde{z}$ with respect 
to a given $(xyz)$ reference system (for example, the one attached to
the unit cell of the crystal). In this case, one should rewrite the ED
Hamiltonian as
\beq
H_{ED}=\sum_{i}\left\{ {\bf E}_S^T \tilde{\chi}^x
{\bf E}_I{\;}\tilde{ M}^x_i+ 
{\bf E}_S^T \tilde{\chi}^y {\bf E}_I{\;}\tilde{ M}^y_i\right\},
\label{H-ED-rotated}
\eeq
where $\tilde{\chi}^{x,y}=R^T \chi^{x,y} R$, 
$\tilde{\bf M}=R  {\bf M}$, and $R$ is the matrix of 
the rotation from $(xyz)$ to $(\tilde{x} \tilde{y} \tilde{z})$. 

%
%
\begin{figure}[htb]
\includegraphics[scale=0.45]{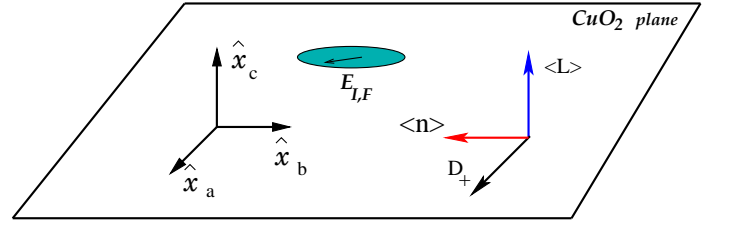}
\caption{Left: LTO coordinate system with respect to a single CuO$_{2}$
  layer. Right: direction of the staggered magnetization, $\langle{\bf
  n}\rangle$, of the DM vector, ${\bf D}_+$, and of the WFM, $\langle{\bf
  L}\rangle\propto\langle{\bf n}\rangle\times{\bf D}_+$, for ${\bf B}=0$. In
  the scattering geometry of \cite{Gozar} the electric field vector of
  the light, ${\bf E}_{I,F}$, is always parallel to the $ab$ plane.}
\label{Fig-1}
\end{figure}
%

\begin{figure}[htb]
\includegraphics[scale=0.45]{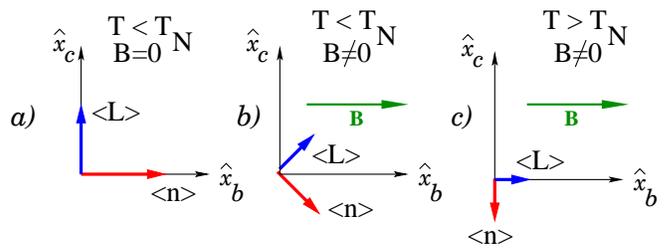}
\caption{Orientation of the staggered magnetization $\langle{\bf n}\rangle$ 
  and of the WFM $\langle{\bf L}\rangle$ for: a) zero field ${\bf B}=0$;
  b) ${\bf B}<{\bf B}_c$ and $T<T_N$ ($\theta$ and $\langle {\bf
  n}\rangle$ are functions of $T$); and c) ${\bf B}<{\bf B}_c$ and
  $T>T_N$. Observe that c) is also realized at any $T$ for ${\bf B}>{\bf B}_c$. }
\label{Fig-2}
\end{figure}

The above Hamiltonian (\ref{H-ED}) is to be treated as a perturbation 
of the following single-layer $S=1/2$ Hamiltonian for the {\lco} system
\beq
H_{S}=J\sum_{\langle i,j\rangle}{\bf S}_{i}\cdot{\bf S}_{j}
+\sum_{\langle i,j\rangle}{\bf D}_{ij}\cdot\left({\bf S}_{i}
\times{\bf S}_{j}\right)+\sum_{\langle i,j\rangle}{\bf S}_{i}
\cdot\overleftrightarrow{\bf \Gamma}_{ij}\cdot{\bf S}_{j},
\label{Hamiltonian}
\eeq
where $J$ is the in-plane antiferromagnetic super-exchange between 
the spins, ${\bf S}_i$, of the neighboring Cu$^{2+}$ ions, and
${\bf D}_{ij}$ and $\overleftrightarrow{\bf \Gamma}_{ij}$ are,
respectively, the DM and XY anisotropic interaction terms of the 
low temperature orthorhombic (LTO) phase of {\lco} \cite{Shekhtman}.
The long-wavelength effective theory for the Hamiltonian (\ref{Hamiltonian}) 
was derived in \cite{MLVC}. As usual, one decomposes the Cu$^{2+}$ spins 
in their staggered, ${\bf n}$, and uniform, ${\bf L}$, components as: 
${\bf S}_i(\tau)/S=e^{\rmi {\bf Q \cdot x}_i} {\bf n}({\bf x}_i,\tau)+ 
{\bf L}({\bf x}_i,\tau)$, where ${\bf Q}=(\pi,\pi)$, and one then 
integrates out ${\bf L}$ to obtain
\bea
{\cal S}&=&\frac{1}{2g_0c_0}\int_{0}^{\beta}\rmd\tau\int\rmd^{2}{\bf x}
\left\{(\partial_{\tau}{\bf n}+\rmi{\bf B}\times{\bf n})^{2}+
c_0^2(\nabla{\bf n})^{2}\right.\nonumber\\
&+&\left.({\bf D_+}\cdot {\bf n})^2+\Gamma_c\; 
(n^c)^2+2 {\bf B}\cdot({\bf D}_{+}\times{\bf n})
\right\},
\label{nlsm}
\eea
where ${\bf n}({\bf x},\tau)$, the continuum analogous of ${\bf M}_i$, is
required to satisfy ${\bf n}^2=1$. Here $g_0$ is the bare coupling
constant, related to the spin-wave velocity, $c_0$, and renormalized
stiffness, $\rho_s$, through $\rho_{s}=c_0(1/Ng_0-\Lambda/4\pi)$
\cite{MLVC}, $\Lambda$ is a cutoff for momentum integrals (we set the
lattice spacing $a=1$), $N=3$ is the number of spin components, ${\bf
D}_+=D_+\hat{\bf x}_a$, and $\Gamma_c>0$. From now one we will use the LTO
($abc$) coordinate system of Fig.\ \ref{Fig-1}, where $\hat{\bf x}_a$, $\hat{\bf x}_b$,
and $\hat{\bf x}_c$ are the LTO unit vectors.  From Eq.\ (\ref{nlsm}) it
follows that the two transverse staggered modes $n^a$ and $n^c$ have their
gaps given by $\Delta_a=D_+=2.5$ meV and $\Delta_c=\sqrt{\Gamma_c}=5$ meV,
respectively, so that $\hat{\bf x}_b$ is the easy-axis for the staggered
magnetization. Since the uniform magnetization $\langle{\bf
L}\rangle=(1/2J)(\langle{\bf n}\rangle \times{\bf D}_{+})$ \cite{MLVC}, we
find that, in the ordered AF phase (for ${\bf B}=0$), a non-zero
weak-ferromagnetic moment (WFM) $\langle{\bf
L}\rangle$ is present, directed along $\hat{\bf x}_c$, see Fig.\ 1 and Fig.\
\ref{Fig-2}(a).

We can now discuss the experiments of \cite{Gozar} in the absence of
magnetic field, ${\bf B}=0$. First, let us observe that since the
sublattice magnetization, $\langle{\bf n}\rangle$, is oriented along the
LTO ${\bf x}_b$ direction, the Hamiltonian (\ref{H-ED}) reads in the continuum
\bea
H_{ED}&=&\int \rmd^{2}{\bf x}\left\{ 
{\bf E}_S^T \chi^a {\bf E}_I{\;}n^a+ 
{\bf E}_S^T \chi^c {\bf E}_I{\;}n^c\right\}\nonumber\\
&=&\int \rmd^{2}{\bf x}
\left\{\Pi_a {\;} n^a+\Pi_c {\;} n^c\right\} ,
\label{H-ED-LTO}
\eea
where we have introduced the polarization projectors 
$\Pi_{a,c}={\bf E}_S^T\chi^{a,c} {\bf E}_I$ with
\bea
\chi^a{=}
  \left( \begin{array}{ccc}
     0 & d & 0 \\
     d^* & 0 & 0\\
     0 & 0 & 0 \\
    
         \end{array} \right)\!\!,{\;\;\;\;\;}
\chi^c{=}
  \left( \begin{array}{ccc}
     0 & 0 & 0 \\
     0 & 0 & d \\
     0 & d^* & 0
         \end{array} \right)\!\!,
\label{matrices}
\eea
in the LTO $abc$ coordinate system.
The structure of the above matrices implies imediately that the $n^a$ (DM)
and $n^c$ (XY) modes should be observed, respectively, in the $B_{1g}$
and $B_{3g}$ channels of the non-magnetic $Dmab$ orthorhombic group of
the LTO phase of {\lco}.

The Brillouin zone center one-magnon Raman intensity can now be 
calculated from Fermi's golden rule using the Hamiltonian 
(\ref{H-ED-LTO}) as a perturbation, and we obtain, for Stokes 
scattering, 
\beq 
{\cal I}(\omega)=[n_B(\omega)+1]\left\{|\Pi_a|^2 {\cal
A}_{a}({\bf 0},\omega)+  |\Pi_c|^2 {\cal
A}_{c}({\bf 0},\omega)\right\},
\label{Raman-intensity}
\eeq
where $n_B(\omega)=(e^{\beta\omega}-1)^{-1}$ is the Bose function and 
${\cal A}_{a,c}({\bf 0},\omega)$ is the ${\bf  q}=0$ spectral function 
of the transverse modes, which is peaked at the mass value $\Delta_{a,c}$.

The scattering geometry used in \cite{Gozar} is the so called {\it
backscattering} geometry. In this setting, the direction of the propagating
wave-vectors of both the incoming and outgoing radiations are chosen to be
along $\hat{\bf x}_c$, in such a way that the electric field of the light
is always parallel to the CuO$_{2}$ or $ab$ plane, see Fig.\
\ref{Fig-1}. Let us denote by $(\hat{e}_{in}\hat{e}_{out})$ the
polarization configurations. Two polarization configurations were used in
\cite{Gozar}: i) light linearly polarized with $\hat{e}_{in}=\hat{\bf x}_a$
and $\hat{e}_{out}=\hat{\bf x}_b$ ($B_{1g}$ geometry); and ii) light 
Right/Left circularly polarized,
$\hat{e}^R=\frac{1}{\sqrt{2}}(\hat{\bf x}_a-\rmi\hat{\bf x}_b)$, and
$\hat{e}^L=\frac{1}{\sqrt{2}}(\hat{\bf x}_a+\rmi\hat{\bf x}_b)$,
combined in the $(RL)$ or $(RR)$ geometries. It is worth mentioning at this
point that the $(RL)$ and $(RR)$ polarization configurations probe,
respectively, the anti-symmetric (imaginary) and symmetric (real) parts of
the Raman tensor $\chi^a$ in (\ref{matrices}), i.e., $\Pi_a^{RL}=2iK_0$ and
$\Pi_a^{RR}=2iG_0S$. Moreover, because the electric-field of the light is
always parallel to the $ab$ plane, $\Pi_c=0$ for any polarization
configuration. As a result, we find that {\em only} the spectral function 
of the $n^a$ mode, ${\cal A}_a({\bf 0},\omega)$, which is peaked at the 
DM gap, $\Delta_a$, contributes to the Raman intensity (\ref{Raman-intensity}) 
in the $B_{1g}$ channel, in agreement with the experiments of \cite{Gozar} 
for both the $B_{1g}$ and $(RL)$ geometries. We observe, furthermore, that 
within the ideal microscopic theory of Fleury and Loudon \cite{Fleury-Loudon}, 
one should always have $G_0=0$ for magnetic systems in which the ground 
state has zero or quenched orbital angular momentum. When this happens, 
also $\Pi^{RR}_a=0$, in which case we could refer to the $(RR)$ geometry 
as {\it forbidden}.

%
%
%

\begin{figure}[htb]
\includegraphics[angle=-90,scale=0.32]{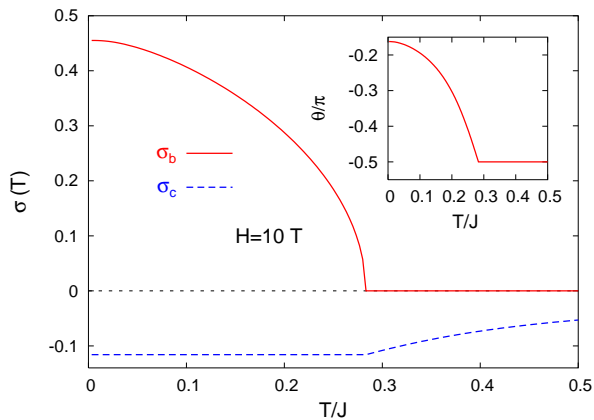}
\caption{$T$ dependence of $\sigma_b$ and $\sigma_c$ at $H=10 T$, from
  Eqs.\ (\ref{T<TN}) and (\ref{T>TN}). Inset: $T$ dependence of the angle
  $\theta$, from Eq.\ (\ref{theta}).}
\label{Fig-3}
\end{figure}

Let us discuss the effects of a magnetic field ${\bf B}$ applied along
$\hat{\bf x}_b$, which is the configuration where the FIM is observed in
Ref. \cite{Gozar}. The first term in Eq.\ (\ref{nlsm}) is responsible for
the softening of the transverse gaps, by an amount $B^2$, $\Delta_{a,c}^2
\rightarrow \Delta_{a,c}^2-B^2$, as observed in the measurements of
Ref. \cite{Gozar}. However, this softening is quantitatively very small for
small fields and will be neglected in what follows.  Conversely, the most
remarkable effect of the magnetic field comes from the last term in Eq.\
(\ref{nlsm}), which can be written as
$(1/g_0c_0)\int_0^\beta\rmd\tau\int\rmd^2{\bf x} {\;}h{\;}n_c$, where
$h=|{\bf B}\times{\bf D}_+|$. As we can see, such term generates, for
$B\parallel\hat{\bf x}_b$, an effective {\em staggered} field $h$ coupled
to $n^c$, leading to a rotation of the staggered-magnetization order
parameter with respect to the zero-field case, see Fig.\ \ref{Fig-2}(b).  
In fact, if we write $\langle{\bf n}\rangle=(0,\sigma_b,\sigma_c)$ for 
the order parameter we obtain at mean-field level
\beq
\label{T<TN}
\sigma_b^2=1-\sigma_c^2-NI_{\perp}(\xi=\infty), \quad
\sigma_c=-h/\Gamma_c,
\eeq
for $T<T_N$ and
\beq
\label{T>TN}
\sigma_b=0, \quad \sigma_c=-h/(\Gamma_c+\xi^{-2}), \quad
1=\sigma_c^2+NI_{\perp}(\xi),
\eeq
for $T>T_N$, where, following the notation of Ref. \cite{MLVC}, $\xi$
is the correlation length and $I_{\perp}=(1/2)(I_{a}+I_{c})$ is the 
integral of the transverse fluctuations \cite{Nota}
$I_{\alpha}=(gT)(2\pi c)\log\left\{(\sinh\left(c\Lambda/2T\right))/
(\sinh\left(\Delta_{\alpha}/2T\right))\right\}$.
The temperature dependence of the components of the 
$\langle{\bf n}\rangle$, as well as of the angle $\theta$,
\beq
\label{theta}
\tan{\theta}=\frac{\sigma_c}{\sigma_b},
\eeq
that defines the rotation of $\langle{\bf n}\rangle$ from the ${\bf x}_b$ 
direction, are reported in Fig.\ \ref{Fig-3}. Observe that for 
$h\neq 0$ we conclude that the uniform magnetization
\beq
\langle{\bf L}\rangle=
\frac{D_+}{2J}\left(|\sigma_b|\hat{\bf x}_c+|\sigma_c|\hat{\bf x}_b\right),
\eeq
is also rotated from its original orientation, perpendicular to the
CuO$_{2}$ planes, by the same angle $\theta$, see Fig. \ref{Fig-2}(b).
Notice, furthermore, that such rotation has the $\hat{\bf x}_a$ axis as
the symmetry axis, in such a way that both $\langle{\bf n}\rangle$ 
and $\langle{\bf L}\rangle$ are always confined to the $bc$ plane 
\cite{Ando-Mag-Anisotropy}. Moreover, the angle of rotation 
{\it varies continuously} from its $T=0$ value until 
$\theta=-\pi/2$, see inset of Fig.\ \ref{Fig-3}, because 
$\sigma_b\rightarrow 0$ {\it continuously} as $T\rightarrow 
T_N(B)$. This is a very important finding because it shows clearly 
that no two-step spin-flop transition occurs, in agreement with
magnetoresistence experiments \cite{Magnetoresistence}. Finally, 
as the field ${\bf B}$ increases the $T=0$ value of $\sigma_b$ 
decreases, until a critical field ${\bf B}_c$ above which the order 
parameter is always oriented along the $\hat{\bf x}_c$ direction. 
As a consequence, either at $T>T_N({\bf B})$, for 
${\bf B}<{\bf B}_c$, or at arbitrary temperature and ${\bf B}>{\bf B}_c$ 
we find a {\it new magnetic state} with $\sigma_b=0$ and 
$\sigma_c\neq 0$, where the WFM is confined to the $ab$ plane 
and ordered along the $\hat{\bf x}_b$ LTO direction, as it was 
suggested by the experiments of \cite{Gozar}. Within our mean-field 
calculation the critical field ${\bf B}_c$ at $T=0$ can be estimated 
from Eq.\ (\ref{T<TN}) as $B_c=\sqrt{1-Ng\Lambda/4\pi}(\Gamma_c/D_+)=40$ T.

%
%
%

We are now ready to explain how the above rotation of the spin easy axis, 
induced by the magnetic field, modifies the Raman spectra. The matrix 
$R$ describing the rotation by the angle $\theta$ around the $\hat{\bf x}_a$
direction 
%
modifies the Raman matrices (\ref{matrices}) to 
\bea
\tilde{\chi}^a&{=}&
  \left( \begin{array}{ccc}
     0 & d\cos\theta & -d\sin\theta \\
     d^* \cos\theta& 0 & 0\\
     -d^*\sin\theta & 0 & 0 \\
         \end{array} \right),\nonumber\\
\tilde{\chi}^c&{=}&
  \left( \begin{array}{ccc}
     0 & 0 & 0 \\
     0 & (d+d^*)\sin\theta\cos\theta & -d^*\sin^2\theta+d\cos^2\theta \\
     0 & -d\sin^2\theta+d^*\cos^2\theta & -(d+d^*)\sin\theta\cos\theta
         \end{array} \right). \nonumber\\
\label{matrices-rotated}
\eea
The new transverse spin modes correspond to $\tilde{n}^a=n^a$, since the
rotation leaves the $\hat{\bf x}_{a}$ direction untouched, and
$\tilde{n}^c=n^c\cos\theta+n^b\sin\theta$. Observe that the main effect of
the rotation is that now $\Pi_c\propto {\bf E}_S^T \tilde{\chi}^{c}{\bf
E}_I\neq 0$, so that also the $n_c$ (XY) mode can be observed. More
specifically, from Eqs. (\ref{H-ED-rotated}) and (\ref{matrices-rotated})
it follows that in the $(RR)$ geometry used in Ref. \cite{Gozar}
$\Pi_a^{RR}=2iG_0\cos\theta$ and $\Pi_c^{RR}= G_0\sin2\theta \cos\theta$. As a
consequence, now also the spectral function of the $n_c$ mode, 
${\cal A}_c({\bf 0},\omega)$, contributes to the intensity
(\ref{Raman-intensity}) with a peak at the energy $\Delta_c$, which 
corresponds to the XY gap. We thus conclude that the FIM mode observed 
in Ref. \cite{Gozar} is nothing less than the XY mode, which can only 
be seen for ${\bf B}\parallel\hat{\bf x}_b$ and nonzero $G_0$. 
%
We should remark at this point that the above effect occurs 
{\it only for} ${\bf B}\parallel\hat{\bf x}_b$, thus justifying the 
appearance of the FIM only for that orientation of the magnetic 
field. In fact, for ${\bf B}\parallel\hat{\bf x}_a$, we have $h=0$ 
and $\sigma_c=0$. In this case, 
$\langle{\bf n}\rangle=\sigma_b\hat{\bf x}_b$, and $\langle{\bf
L}\rangle=(1/2J)D_+\sigma_b\hat{\bf x}_c$, see Fig.\ \ref{Fig-2}(a). The
case of ${\bf B}\parallel\hat{\bf x}_c$ was already discussed in
\cite{MLVC}. Although in this third case $h\neq 0$, it is actually a source
for the $n_b$ component of the staggered magnetization, and we again have
$\sigma_c=0$. Thus, for either ${\bf B}\parallel\hat{\bf x}_a$ or ${\bf
B}\parallel \hat{\bf x}_c$, where $\theta=0$, we will have $\Pi_c=0$ and
only the DM mode will be present in the one-magnon Raman spectrum.

In conclusion, we have found that the appearance of the FIM for 
${\bf B}\parallel{\bf x}_b$ is a consequence of both a 
{\it continuous} rotation of the spin-quantization basis and 
of a nonzero symmetric part for the $\chi^c$ Raman tensor, 
$G_0\neq 0$, signaling a deviation from the ideal ED scattering 
mechanism of Fleury and Loudon \cite{Fleury-Loudon}. Furthermore, 
we have found a {\it new magnetic state}, well above $T_N$, in 
which the WFM is confined to the $ab$ plane. The absolute intensities
of the one-magnon Raman peaks in the two circular polarizations,
$(RL)$  and $(RR)$, are given in terms of $K_0$ and $G_0$, which
have to be determined from first principles. Moreover, to obtain the
temperature and field dependencies of the relative peak intensity of
the FIM a more precise calculation of the $T$ and $B$ dependencies of
the anisotropy gaps and of the spin damping would be required.

The authors have benefitted from invaluable discussions with Y.~Ando, 
G.~Blumberg, S.~Caprara, A.~Gozar, M.~Grilli, A.~N.~Lavrov, and 
J.~Lorenzana.


\begin{thebibliography}{20}
\bibitem{Nota}
We fix $\rho_s=0.1J$, $c_0=1.3J$ and $J=100$ meV by fitting 
the experimental data for the correlation length $\xi$, with
Eq.\ (\ref{T>TN}), see Ref.\ \cite{MLVC}.
\end{thebibliography}
\end{document}